\documentclass[%
 reprint,
 floatfix,
 amsmath,
 amssymb,
 aps,
 prl,
]{revtex4-2}

\usepackage{graphicx}
\usepackage{dcolumn}
\usepackage{bm}

\usepackage{color} 
 
\newcommand{\ve}[1]{\boldsymbol{#1}}

\newcommand{\ydnote}[1]{\textcolor{blue}{#1}} 

\begin{document}

\preprint{APS/123-QED}

\title{Autonomous navigation of shape-shifting microswimmers}

\author{Yong Dou}%
\author{Kyle J. M. Bishop}%
 \email{kyle.bishop@columbia.edu}
\affiliation{%
 Department of Chemical Engineering, Columbia University, New York, NY 10025 USA
}%

\date{\today}

\begin{abstract}
    We describe a method for programming the autonomous navigation of active colloidal particles in response to spatial gradients in a scalar stimulus.  Functional behaviors such as positive or negative chemotaxis are encoded in the particle shape, which responds to the local stimulus and directs self-propelled particle motions.  We demonstrate this approach using a physical model of stimuli-responsive clusters of self-phoretic spheres.  We show how multiple autonomous behaviors can be achieved by designing the particle geometry and its stimulus response.
\end{abstract}

\maketitle

\begin{figure}[h!]
     \centering
     \includegraphics{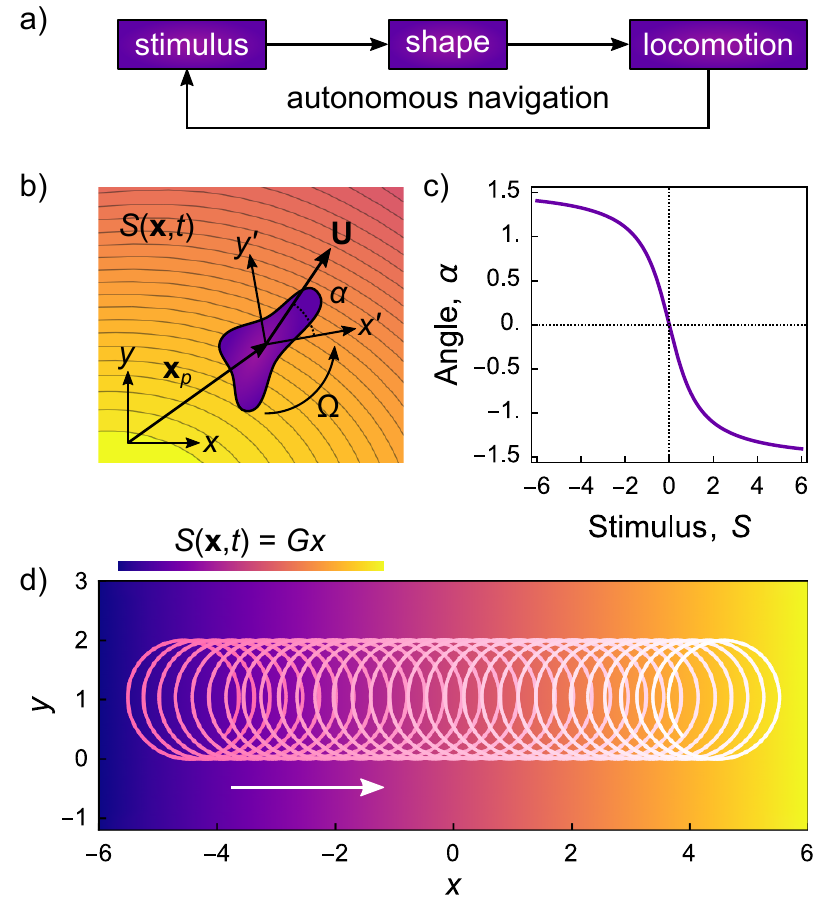}
     \caption{
     (a) A local stimulus determines particle shape which in turn directs particle motion towards new regions of the stimulus landscape.
     (b) An active particle moves with linear velocity $\ve{U}= U\cos\alpha \ve{e}_{x'} + U\sin\alpha \ve{e}_{y'}$ and angular velocity $\Omega\ve{e}_{z'}$ across a two-dimensional stimulus landscape $S(\ve{x},t)$.  
     (c) Assuming $R=U/\Omega=\text{constant}$, particle trajectories are determined by the stimulus-dependent orientation of the propulsion velocity $\alpha(S)$. Here, the function $\alpha(S)$ encodes for positive chemotaxis---that is, autonomous motion toward regions of higher stimulus. 
     (d) Computed trajectory of the particle in (c) on a uniform stimulus gradient $S(\ve{x})=G x$ of magnitude $G=0.1$. Here, lengths are scaled by the curvature radius $R$.}
     \label{fig:1}
 \end{figure}

\paragraph*{Introduction.} 
Chemotactic bacteria swim autonomously through complex media to regulate their environment and find new energy sources. Physically, these functional behaviors are enabled by feedback bet¬ween local sensing and autonomous motion across stimulus landscapes that vary in space and time (Fig.\ \ref{fig:1}a). Depending on the relationship between sensing and motion, different functional behaviors can be achieved (e.g., positive or negative taxis, kinesis). In particular, bacteria use temporal sampling and biochemical memory to bias their run-and-tumble motions in even weak gradients, which cannot be detected directly over the length of the organism \cite{cates2012diffusive}.  

By contrast, attempts to mimic such autonomous propulsion and navigation in synthetic colloids \cite{bechinger2016active} have relied on particle alignment within stimulus gradients to direct particle motion (e.g., gradients of chemical concentration \cite{hong2007chemotaxis,popescu2018chemotaxis}, magnetic potential \cite{Kline2005}, light intensity \cite{lozano2016phototaxis}, fluid velocity \cite{Palacci2015}, and fluid viscosicty \cite{liebchen2018viscotaxis}).  This approach does not scale favorably to micron-scale colloids moving in weak gradients.  Moreover, the mechanisms of self-propulsion (e.g., self-phoresis \cite{golestanian2007designing,popescu2018chemotaxis}) are often re-purposed for navigation, which presents challenges for designing different functional behaviors such as positive \emph{or} negative chemotaxis.  The realization of colloidal robots \cite{palagi2018bioinspired,han2018engineering} that navigate autonomously through fluid environments requires new strategies for programming active particles to bias their motion in response to local stimuli.

Here, we propose one such strategy based on the shape-directed propulsion of shape-shifting microswimmers that alter their shape and thereby their motion in response to changes in a scalar stimulus (e.g., the concentration of a particular chemical species). Particle shape provides a versatile medium for encoding the dynamical behavior of active colloids powered by a variety of energy inputs (e.g., electric \cite{brooks2018shape}, acoustic \cite{sabrina2018shape}, self-electrophoretic \cite{brooks2019shape}). Moreover, by using stimuli-responsive materials, microscale particles can be designed to change their shape in response to environmental cues \cite{magdanz2014stimuli,palagi2016structured}.  We describe how these two concepts can be integrated to design active colloids that navigate autonomously across heterogeneous stimulus landscapes.   

We consider a single self-propelled particle moving on a two-dimensional domain with linear velocity $\ve{U}=U\cos\alpha \ve{e}_{x'} + U\sin\alpha \ve{e}_{y'}$ and angular velocity $\ve{\Omega}=\Omega \ve{e}_{z'}$ (Fig.\ \ref{fig:1}a). In the particle frame of reference, the velocity components---parameterized by $U$, $\alpha$, and $\Omega$---depend only on an internal state variable $s\in[0,1]$, which describes a single degree of freedom in the particle shape.  The internal state of the particle depends in turn on the local value of a scalar stimulus field $S(\ve{x},t)$ such as the concentration of a chemo-attractant or repellent.  We assume that the state variable $s$ evolves rapidly to changes in the stimulus and is uniquely determined by its magnitude at the particle center $\ve{x}_p$---that is, $s=f(S(\ve{x}_p,t))$. For a given stimulus landscape $S(\ve{x},t)$, the dynamics of the particle is therefore determined by the response functions, $U(S)$, $\alpha(S)$, and $\Omega(S)$, which describe how particle motion depends on the magnitude of the local stimulus.

\paragraph*{Particle motion in stimulus gradients.} 
The response functions can be designed to enable particle migration up (or down) spatial gradients in the stimulus landscape. In a homogeneous environment, the particle moves with a constant speed along circular orbits of radius $R=U/\Omega$, larger than the characteristic size of the particle $L$.  Although the particle cannot detect stimulus gradients directly, it can integrate the effects of such gradients over each circular orbit to produce steady motions guided by the gradient. 

For example, a uniform gradient in the $x$-direction, $S(\ve{x})=G x$, causes the particle to drift with velocity $\ve{V}=-\tfrac{1}{2}G U R \alpha' \ve{e}_x + \tfrac{1}{2}G U R' \ve{e}_y + \mathcal{O}(G^2)$, where primes denote differentiation with respect to the stimulus $S$ \cite{Supp}. To prohibit motion perpendicular to the gradient direction, the radius $R$ of the particle trajectory should be designed to be independent of the stimulus magnitude. Higher order contributions to the drift velocity are negligible provided that drift is much slower than propulsion---that is, when $V\ll U$ or, equivalently, when $G\ll 2/R\alpha'$. With these assumptions, the resulting particle trajectory depends only on the response function $\alpha(S)$, which characterizes the orientation of the propulsion velocity in the particle frame.

Fig.\ \ref{fig:1}d shows one particle trajectory for the response function $\alpha(S) = -\arctan[(S-S_l)/S_s]$, where $S_l$ and $S_s$ are location and scale parameters. Without loss of generality, we set these parameters to zero and one, respectively, such that all stimuli are given in dimensionless form ($S_l\rightarrow0$ and $S_s\rightarrow1$).  This response acts to rotate the propulsion velocity in the particle frame by up to 180$^{\circ}$ as the local stimulus changes.  The drift speed reaches a maximum value of $V_{\max}=G U R$ at $S=0$ and decays as $V=G U R / S^2$ for extreme stimulus magnitudes ($\lvert S \rvert \gg 1$).

\paragraph*{Self-phoretic clusters.}  
In practice, the design of the response functions $U(S)$, $\alpha(S)$, and $\Omega(S)$ requires one to consider the specific mechanism(s) of particle propulsion and its dependence on particle shape. Here, we consider the self-phoretic propulsion of hard sphere clusters, which have been studied previously in theory \cite{soto2014self, varma2018clustering} and experiment \cite{niu2018dynamics, schmidt2019light}. Each sphere $j$ in the cluster emits a constant flux $A_j$ of some chemical species, which sets up a concentration field $c(\ve{x})$ around the composite particle. At small P\'eclet number ($\text{Pe}=UL/D\ll1$), the species concentration is governed by the Laplace equation for steady-state diffusion, $\nabla^2c=0$, subject to the following boundary condition on the surface $\mathcal{S}_j$ of each sphere $j$
\begin{equation}
   -D\ve{n} \cdot \nabla c(\ve{x}) = A_j \quad \text{for} \quad \ve{x}\in \mathcal{S}_j
\end{equation}
where $D$ is the species diffusivity, and $\ve{n}$ is the unit normal directed out from the spheres.  Far from the particle, the species concentration approaches a constant value $c^{\infty}$, which can be set to zero without loss of generality. The resulting concentration field $c(\ve{x})$ depends on the shape of particle---that is, the configuration of its component spheres---but not its position and orientation within the stimulus landscape.

Concentration gradients tangent to the surface of each sphere drive interfacial phoretic flows with velocity 
\begin{equation}
    \ve{u}(\ve{x})=-\mu_j (\ve{\delta}-\ve{n}\ve{n}) \cdot \nabla c\quad \text{for} \quad \ve{x}\in \mathcal{S}_j
\end{equation}
where  $\mu_j$ is the mobility coefficient of sphere $j$.  At low Reynolds number ($\text{Re}=\rho UL/\eta$), the resulting velocity and pressure fields are governed by the Stokes equations, $-\nabla p + \eta \nabla^2 \ve{u}=0$ and $\nabla \cdot \ve{u}=0$, where $\eta$ is the fluid viscosity.  The linear velocity $\ve{U}$ and angular velocity $\ve{\Omega}$ of the rigid cluster is determined by the condition that there is no net force or torque on the particle.  We use a far-field approximation based on the method of reflections \cite{varma2018clustering} to solve both the diffusion and hydrodynamic problems outlined above and estimate the particle velocity as a function of its shape \cite{Supp}. For simplicity, we focus on the specific case of homogeneous clusters with $A_j=A$ and $\mu_j=\mu$; however, the model can also describe heterogeneous clusters made from spheres of different types.  We scale lengths by $L$, concentrations by $A L/D$, and velocities by $\mu A/D$, such that the velocities $\ve{U}$ and $\ve{\Omega}$ are determined entirely by particle geometry.  

\begin{figure}[!h]
    \centering
    \includegraphics[width=8.5cm]{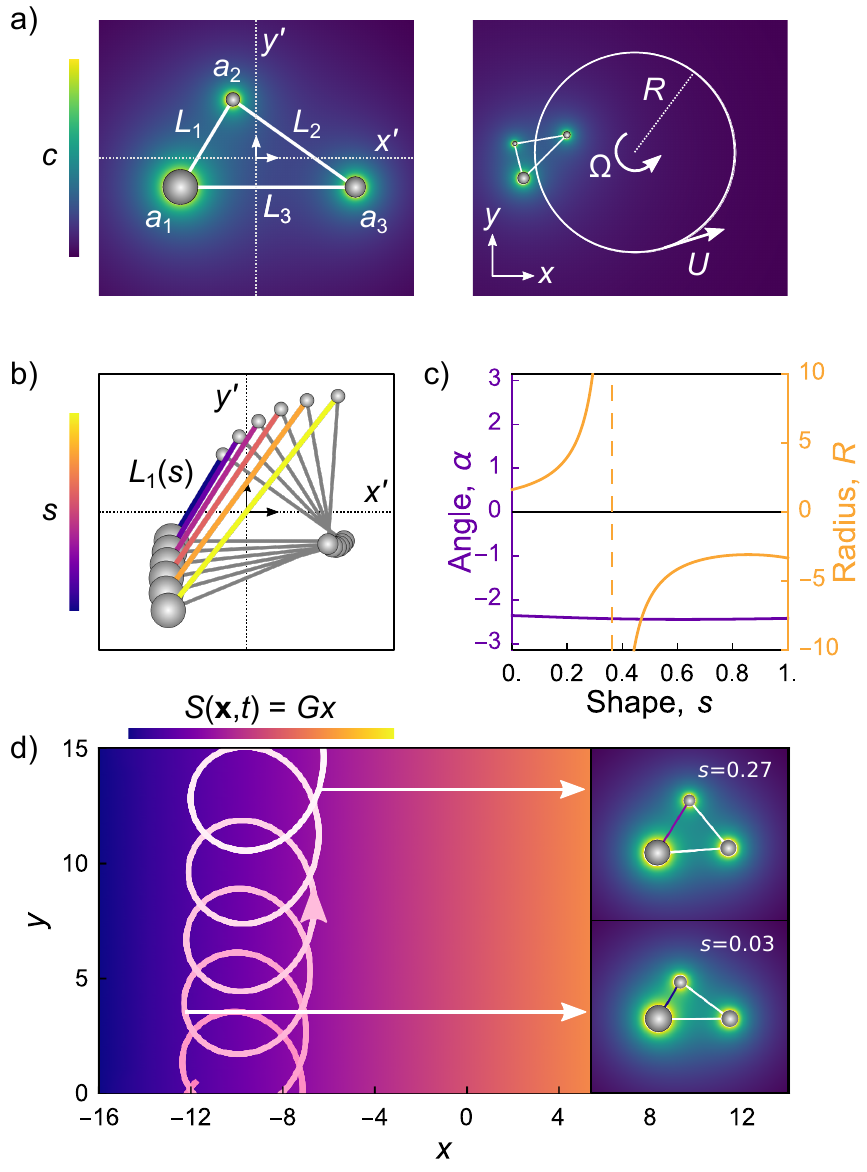}
    \caption{Shape-shifting self-phoretic clusters. (a) Spheres connected by rigid bonds emit a chemical species that drives phoretic flows and particle motion with linear velocity $\ve{U}$ and angular velocity $\Omega$. Rigid particles move along circular trajectories of (signed) radius $R=U/\Omega$ (right). 
    (b) Standard shapes for different values of the shape parameter $s=(L_1-L_{\min})/(L_{\max}-L_{\min})$, which controls the length of one bond. The sphere positions $\ve{x}'_j(s)$ in the particle frame are chosen to eliminate particle translation and rotation due to shape change. 
    (c) Response functions $R(s)=U(s)/\Omega(s)$ and $\alpha(s)$ for the standard shapes in (b). 
    (d) Computed particle trajectory on a uniform stimulus gradient $S(\ve{x})=Gx$ of magnitude $G=0.2$; the shape parameter is assume to vary with the local stimulus as $s = (1+ e^{-S})^{-1}$. 
    In (a-d), the sphere radii are $a_1=0.1$, $a_2=0.04$, $a_3=0.06$; the fixed bond lengths are $L_2=0.860$ and $L_3=1$; the length of the responsive bond $L_1$ varies from $L_{\min}=0.5$ to $L_{\max}=1.5$.}
     \label{fig:2}
 \end{figure}

\paragraph*{Shape-shifting clusters.}  
In addition to shape-directed particle motion, we must also consider the effects of shape-shifting whereby the particle shape changes in response to changes in the local stimulus $S(\ve{x}_p,t)$.  Such particles can now be realized in experiment using stimuli responsive soft materials such as shape-changing polymers \cite{magdanz2014stimuli} or liquid crystal elastomers \cite{palagi2016structured} to alter the sizes and/or relative positions of spheres in the cluster.  In the present model, we consider a highly idealized form of shape-shifting, in which the bond lengths between neighboring spheres can depend on the local stimulus.  Fig.\ \ref{fig:2}b shows one example of a three-sphere cluster containing one stimuli-responsive bond.

To describe the motion of shape-shifting particles, we define a set of standard shapes that specify the position $\ve{x}_j'(s)$ of each sphere in the particle frame as a function of the shape parameter $s$ \cite{shapere1989geometry}.  The standard shapes are chosen such that the particle frame does not translate or rotate within the viscous fluid as the particle changes its shape \cite{Supp}.  In this way, we effectively remove the effects of particle swimming due to shape change, allowing us to focus exclusively on shape-induced changes in the propulsion velocity.  

For each standard shape, we compute the linear velocity $\ve{U}$ and angular velocity $\ve{\Omega}$ of the (rigid) cluster as described above to obtain the response functions $U(s)$, $\alpha(s)$, and $\Omega(s)$. Fig.\ \ref{fig:2}c shows the response functions for a triangular cluster as a function of its one stimuli-responsive bond length, $s = (L_1 - L_{\min})/(L_{\max}-L_{\min})$.  This particle exhibits drifting motions in a stimulus gradient, but not the desired chemotactic motions parallel to the gradient direction.  In general, clusters will not satisfy the design criterion for chemotaxis that $R=\text{constant}$; however, it is possible to optimize their response by modifying the particle geometry.

\paragraph*{Design of chemotactic clusters.} In designing chemotactic clusters that navigate up (or down) stimulus gradients, we seek to alter the cluster geometry such that each of the standard shapes leads to self-phoretic motion with a constant (signed) radius $R_0$. The design process can be formulated as an optimization problem that seeks to minimize the objective
\begin{equation}
    O(\ve{d}) = \langle [R(s,\ve{d})-R_0]^2 \rangle_s
\end{equation}
Here, the cluster geometry is parameterized by both the shape parameter $s\in[0,1]$ and the design variable $\ve{d}$, which is held fixed during shape-shifting.  The angle brackets denote averages over the shape parameter $s$, and $R(s,\ve{d})$ is the computed radius of the particle trajectory.

For the three-sphere cluster in Fig.\ \ref{fig:2}a, the design variable $\ve{d}$ includes two of the three radii ($a_2,a_3$) and one of the three bond lengths ($L_2$). All lengths are scaled by the length of the third bond $L_3$ such that $L_3\rightarrow 1$; the sphere radius $a_1$ is specified such that $a_1\ll L$. The length $L_1$ of the stimulus-responsive bond is constrained to vary between user-specified limits $L_{\min}$ and $L_{\max}$. For each design $\ve{d}$, we compute the radius $R(s,\ve{d})$ of the particle trajectory as a function of the shape parameter using the model outlined above.  Numerical optimization methods can then be applied to identify the optimal shape-shifting particle.

\begin{figure}[!h]
    \centering
    \includegraphics[width=8.5cm]{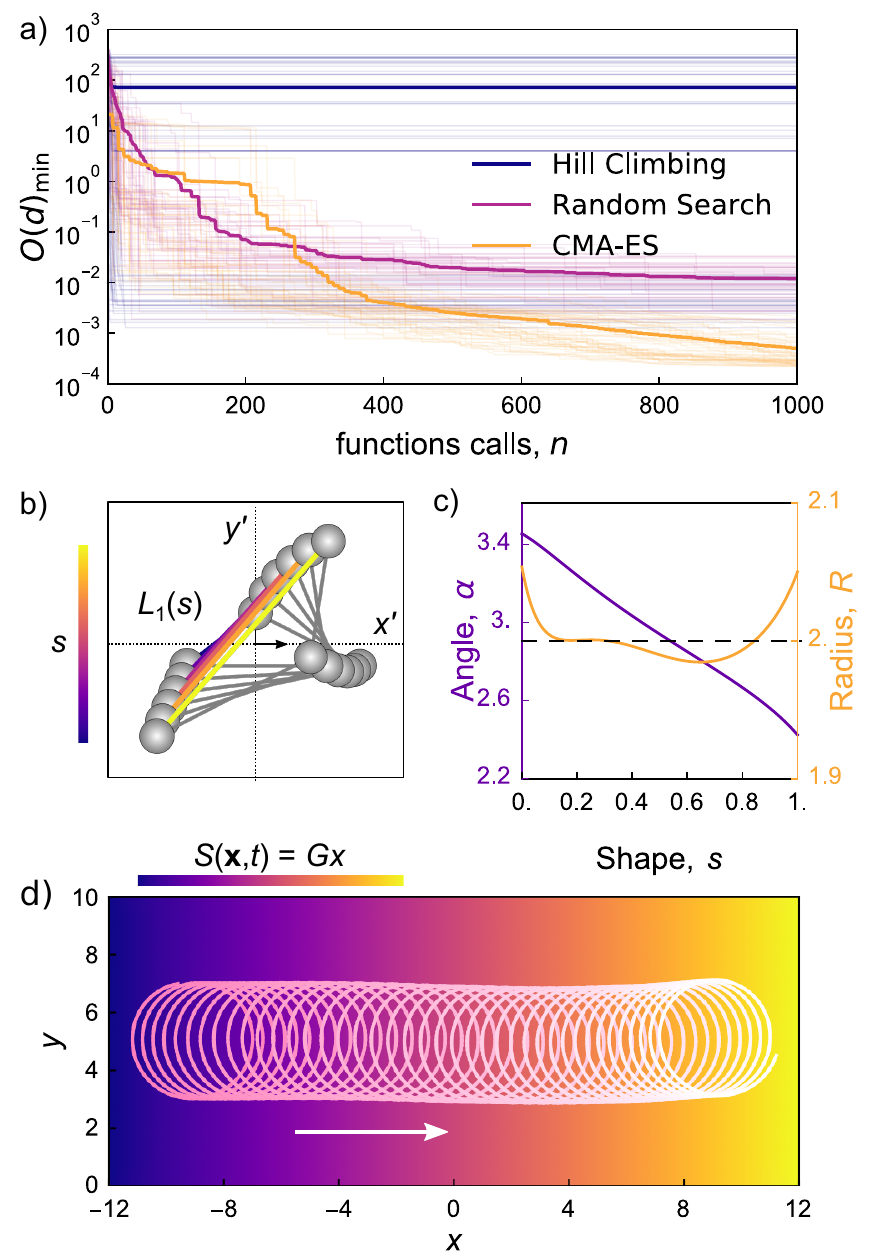}
    \caption{Design of a chemotactic three-sphere cluster. (a) Minimum objective function $O(\ve{d})$ identified after $n$ function evaluations for three optimization methods initialized from 50 randomly selected designs (light curves); bold curves show the average performance. (b) Standard shapes for the optimal design for different values of the shape parameter $s=(L_1-L_{\min})/(L_{\max}-L_{\min})\in [0,1]$. During optimization, we prescribe the lengths $L_3=1$, $L_{\min}=0.5$, $L_{\max}=1.5$, $a_1=0.1$, and $R_0=2$; the optimal design identified is $a_2=0.0985$, $a_3=0.0958$, and $L_2=0.677$. (c) Computed response functions $\alpha(s)$ and $R(s)=U(s)/\Omega(s)$ for the standard shapes in (b). (d) Computed particle trajectory on a uniform stimulus gradient $S(\ve{x})=Gx$ of magnitude $G=0.2$; the shape parameter is assume to vary with the local stimulus as $s = (1+ e^{-S})^{-1}$.}
    \label{fig:3}
\end{figure}

Fig.\ \ref{fig:3}a highlights the performance of three different optimization methods based on hill climbing, random search, and the covariance matrix adaptation evolution strategy (CMA-ES) \cite{hansen2016cma}, as applied to the design of chemotactic three-sphere clusters.  Greedy search algorithms such as hill climbing are quickly trapped in local minima; random searches are more effective but fail to identify the deepest minima.  The CMA-ES algorithm, which has proven effective on other problems involving colloidal clusters \cite{miskin2013adapting}, combines stochastic ``mutation'' events with a deterministic ``selection'' process to identify optimal cluster geometries.  Fig.\ \ref{fig:3}b shows the optimal design for a three-sphere cluster with one stimuli-responsive bond.  Variations in the radius $R$ are ca.\ 2\% of the prescribed value (Fig.\ \ref{fig:3}c).  The orientation of the propulsion velocity $\alpha(s)$ decreases almost linearly with the shape parameter $s$ over a range spanning ca.\ 60$^{\circ}$. Assuming a sigmoidal response of the shape parameter on the local stimulus, the designed particle moves steadily up an applied stimulus gradient (positive chemotaxis; Fig.\ \ref{fig:3}d).

For spheres with the same activity and mobility connected by the same type of stimuli responsive bonds, one can design other particle clusters that behave in the opposite manner and swim down an applied gradient \cite{Supp}.  Moreover, by altering the objective function, one can design particle clusters that navigate perpendicular to the gradient in a specified direction \cite{Supp}. In addition to shape-shifting bonds between rigid spheres, one can also use responsive spheres that swell or shrink in response to local conditions. More generally, the present navigation strategy can be applied whenever particle motion depends on the local stimulus magnitude. Particle shape provides one of several possible design strategies for mediating this relationship between stimulus and motion.

\paragraph*{Effects of noise.} 
The type of autonomous navigation described here is only effective when rotational diffusion is slower than self-phoretic particle rotation.  Accounting for the particle's Brownian motion, the drift velocity in a uniform stimulus gradient $S(\ve{x},t)=G x$ can be approximated as $\ve{V}=-\tfrac{1}{2} G U R \alpha' \text{Pe}^2_r / (1+ \text{Pe}_r^2) \ve{e}_x +\mathcal{O}(G^2)$, where $\text{Pe}_r =\Omega\xi_r/k_B T$ is a rotational P\'eclet number, $\xi_r$ is a rotational friction coefficient (e.g., $\xi_r=8\pi\eta a^3$ for a sphere in an unbounded fluid), and $k_B T$ is the thermal energy.  As detailed in \cite{Supp}, the derivation of this expression assumes that both the linear and angular propulsion velocity are independent of the stimulus (i.e., $U(S)=U$ and $\Omega(S)=\Omega$) and that the angular response function $\alpha(S)$ is approximately linear over changes in stimulus of order $GR$.  

For large P\'eclet numbers ($\text{Pe}_r\gg1$), the particle maintains orientational correlations over the duration of each circular orbit and uses this information to guide its motion on the stimulus landscape.  By contrast, rotational diffusion at small P\'eclet numbers ($\text{Pe}_r\ll1$) acts to erase these correlations thereby prohibiting effective navigation. Approximating the particle as a thin disk of diameter $L$ (such that $\xi_r=\tfrac{4}{3}\eta L^3$ \cite{Kim2005}), the minimum particle size required for effective navigation ($\text{Pe}_r\sim1$) is $L\sim(3k_B T/4\Omega\eta)^{1/3}$.  Assuming a typical phoretic velocity of $\Omega=1$ rad/s, autonomous navigation in water ($\eta=10^{-3}$ Pa s) at room temperature requires particles sizes of order $L\sim 1~\mu$m or larger. Perhaps not surprisingly, this length is comparable to that of chemotactic bacteria such as \emph{E. coli}.

\paragraph*{Conclusions.}  In sum, the shape-directed dynamics of shape-shifting particles provides a viable strategy for designing active colloids capable of autonomous navigation through heterogeneous environments.  While the present model focuses on self-phoretic motion, similar strategies should apply to other propulsion mechanisms that depend on particle shape (e.g., those based on external electric \cite{Ma2015} or magnetic \cite{Driscoll2017} fields).  The design of desired responses requires predictive understanding of both the relationship between stimulus and shape as well as that between shape and motion.  This design problem is simplest when the processes of shape-shifting and of shape-directed motion are effectively decoupled (e.g., when the former is much faster than the latter as assumed here). Recent advances in combining active colloids and stimuli-responsive materials should provide a promising platform to apply the design principles outlined here \cite{Alvarez2019}. Looking forward, autonomous navigation based on internal degrees of freedom could be combined with external control strategies \cite{liebchen2019optimal} to create colloidal robots capable of increasingly intelligent functional behaviors.

\paragraph*{Acknowledgements.} This work was supported by the Center for Bio-Inspired Energy Science, an Energy Frontier Research Center funded by the U.S. Department of Energy, Office of Science, Basic
Energy Sciences under Award DE-SC0000989. 

\bibliography{references}

\end{document}